\documentclass[structabstract]{aa}  
\usepackage{graphicx}
\usepackage{txfonts}
\usepackage{natbib}
\bibpunct{(}{)}{;}{a}{}{,} 
\begin{document}
\newcommand{\xmm} {\sl XMM-Newton}
\newcommand{\ergscm} {erg s$^{-1}$ cm$^{-2}$}
\newcommand{\ergss} {erg s$^{-1}$}
\newcommand{\ergsd} {erg s$^{-1}$ $d^{2}_{100}$}

   \title{Phase-resolved optical and X-ray spectroscopy of low-mass X-ray binary X1822--371\thanks{Based on observations obtained at the European Southern Observatory, Chile (Program-ID: 077.D-0603(A)) and with XMM-Newton, an ESA
science mission with instruments and contributions directly funded by ESA Member States and NASA.}
   	}

   \subtitle{}

   \author{A. Somero
          \inst{1}
          \and
          P. Hakala\inst{2}
          \and
          P. Muhli\inst{3}
          \and
          P. Charles\inst{4,}\inst{5,}\inst{6}
          \and
          O. Vilhu\inst{7}
          }

	\institute{Tuorla Observatory, Department of Physics and Astronomy, University of Turku, V\"{a}is\"{a}l\"{a}ntie 20, FI-21500 Piikki\"{o}, Finland\\
              \email{aunsom@utu.fi}
        \and
         Finnish Centre for Astronomy with ESO (FINCA), University of Turku, V\"{a}is\"{a}l\"{a}ntie 20, FI-21500 Piikki\"{o}, Finland
        \and
	National Land Survey of Finland, Opastinsilta 12C, PL 84, FI-00521 Helsinki, Finland
        \and
         School of Physics \& Astronomy, University of Southampton, Southampton SO17 1BJ, UK
	\and
	South African Astronomical Observatory, P.O. Box 9, Observatory 7935, South Africa
	\and
	Department of Astronomy, University of Cape Town, Cape Town, South Africa
	\and
	Division of Geophysics and Astronomy, Department of Physics, University of Helsinki, P.O.Box 64, FI-00014 Helsinki, Finland
             }

   \date{Received 11 November 2011 / Accepted 25 December 2011}

 
  \abstract
   {X1822--371 is the prototypical accretion disc corona X-ray source, a low-mass X-ray binary viewed at very high inclination, thereby allowing the disc structure and extended disc coronal regions to be visible.
   As the brightest (closest) such source, X1822--371 is ideal for studying the shape and edge structure of an accretion disc, and comparing with detailed models.
   }
   {We study the structure of the accretion disc in X1822--371 by modelling the phase-resolved spectra both in optical and X-ray regime.   
   }
   {We analyse high time resolution optical ESO/VLT spectra of X1822--371 to study the variability in the emission line profiles. 
   In addition, we use data from \textit{XMM-Newton} space observatory to study phase-resolved as well as high resolution X-ray spectra.
   We apply the Doppler tomography technique to reconstruct a map of the optical emission distribution in the system.
   We fit multi-component models to the X-ray spectra.
   }
   {We find that our results from both the optical and X-ray analysis can be explained with a model where the accretion disc has a thick rim in the region where the accretion stream impacts the disc. 
   The behaviour of the H$\beta$ line complex implies that some of the accreting matter creates an outburst around the accretion stream impact location and that the resulting outflow of matter moves both away from the accretion disc and towards the centre of the disc. 
   Such behaviour can be explained by an almost isotropic outflow of matter from the accretion stream impact region.
The optical emission lines of \ion{He}{ii} $\lambda$4686 and 5411 show double peaked profiles, typical for an accretion disc at high inclination. 
However, their velocities are slower than expected for an accretion disc in a system like X1822--371. 
This, combined with the fact that the \ion{He}{ii} emission lines do not get eclipsed during the partial eclipse in the continuum, suggests that the line emission does not originate in the orbital plane and is more likely to come from above the accretion disc, for example the accretion disc wind.
    }
   {}

\authorrunning{A. Somero et al.}
\titlerunning{Phase-resolved optical and X-ray spectroscopy of LMXB X1822--371}

\maketitle
%

\section{Introduction} \label{Intro}
\object{X1822--371} (\object{V691 CrA}) is a low-mass X-ray binary (LMXB) system, consisting of a neutron star (NS) primary that accretes matter from a Roche-lobe filling low-mass secondary star.
The nature of the primary has been confirmed to be a pulsar with the discovery of 0.59 s X-rays pulsations \citep{jonker}.
The orbital period of the system is 5.57 hr and it is seen over a wide range of wavelengths: X-rays \citep{white}, UV \citep{puchnarewicz}, optical \citep{mason} and infrared \citep{masoncordova}.
The basic concept of emission from LMXBs is that the X-rays are produced by the accretion process and the less energetic radiation through the reprocessing of the X-rays in the accretion disc and the secondary star.
The orbital inclination of the system has been determined to be $i = 82.5 \degr \pm 1.5\degr $ \citep{heinz}.
The partial nature of the X-ray eclipse by the secondary star has been explained with the presence of an accretion disc corona (ADC), an extended cloud of hot ionized gas surrounding the NS and an accretion disc which is big enough so that the secondary cannot totally block it from our line of sight.
The ADC also explains why the ratio of X-ray to optical luminosity in X1822--371 is atypically low ($L_\mathrm{X}/L_\mathrm{opt} \approx 20$ \citep{griffiths}, usually it is at least one order of magnitude larger for LMXBs).
This is because the ADC is thought to be optically thick in X-rays and thus we cannot see the central source and observe only scattered X-rays.

The accretion disc in X1822--371 is considered to be thick and have non-axisymmetric vertical structure.
Variability of the light curve outside the eclipse is explained to be due to the extended accretion disc rim \citep{masoncordova, helliermason} blocking our view of the inner parts of the accretion disc.
From the X-ray pulse arrival times \citet{jonker} were able to determine that the orbit of the system is almost circular and that the mass function of the neutron star is $(2.03 \pm 0.03) \times 10^{-2} M_{\sun}$.
\citet{casares} detected Bowen fluorescence from the secondary star and constrained the lower limit of the radial velocity semiamplitude of the secondary to be $300 \pm 8$ km s$^{-1}$.
This combined with the radial velocity semiamplitude of the neutron star (calculated from the projected semi-major axis of the NS by \citet{jonker}) constrains the lower limits for the component masses $M_2>0.36 M_{\sun}$ and $M_1>1.14 M_{\sun}$ \citep{casares}.
However, \citet{munoz} take the analysis further by estimating the correction for non-uniform  irradiation of the secondary star, which leads to a lower limit of $M_1>1.6M_{\sun}$.
The orbital period of X1822--371 has been observed to be increasing \citep{hellier90, bayless, burderi}, which has been explained by the presence of a wind or outflow carrying angular momentum away from the system.

We have organized this paper in the following way.
Section 2 describes our optical observations and data reduction.
In Section 3 we present analysis of the optical data.
In Section 4 we present X-ray observations, data reduction and analysis.
Finally in Section 5 we discuss our results and present our conclusions.


\section{Observations and data reduction}

We observed X1822--371 with the Focal Reducer/Low Dispersion Spectrograph (FORS) 2 instrument \citep{appenzeller} at the Very Large Telescope (VLT) of the European Southern Observatory (ESO) on 15 June 2006.
The detector of FORS2 consists of two 4096$\times$2048 pixel MIT CCDs (15$\mu$m/pixel) and they were read out in 2$\times$2 binning mode \citep{forsmanual}.

The data were obtained with grism 1400V (4560--5860 \AA, dispersion 20.8 \AA\ mm$^{-1}$) and a 0.7\arcsec slit.
The slit was oriented so that it included a nearby star $\sim$6.7\arcsec south of X1822--371 for comparison in order to monitor slit losses and possible atmospherical changes. 
The exposure time of a single spectrum was 90 seconds which, adding the overheads, yielded a time resolution of about 2 minutes.
In total we obtained 224 spectra covering approximately 1.6 orbital cycles of X1822--371.
The seeing varied between 0.49\arcsec and 1.41\arcsec during the night, and averaged 0.87\arcsec.

The calibration frames were obtained as part of the standard FORS Calibration Plan provided by ESO, these included bias frames, screen spectroscopic flat field images and an arc lamp spectrum all taken in the morning after the observations.
The data reduction was conducted by using standard IRAF\footnote{IRAF is distributed by the National Optical Astronomy Observatories, which are operated by the Association of Universities for Research in Astronomy, Inc., under cooperative agreement with the National Science Foundation.}/PyRAF\footnote{PyRAF is a product of the Space Telescope Science Institute, which is operated by AURA for NASA.} packages for debiasing, flat-fielding, extracting the spectra and wavelength calibration.

Arc lamp spectra were extracted with the same apertures as used for the spectra of X1822--371 and the comparison star.
The wavelength solution was found by fitting 7 lines with a first order cubic spline yielding an RMS of 0.02-0.03 \AA.
The dispersion of the data is $\sim 0.64$ \AA /pixel and the resolution, measured from the arc lamp and the night sky [\ion{O}{i}] lines, is $\sim 1.4$ \AA.
As the ESO calibration plan provides only one arc line exposure, we used the position of the sky [\ion{O}{i}] emission line at 5577$ \AA$ to check the wavelength calibration zero point.
The line was fitted with a Gaussian profile and shifts were calculated as the deviation of the line centre from 5577.34 \AA\ for both the apertures of X1822--371 and the comparison star separately.
The shifts were found to be between $-0.465$ \AA\ and 0.345 \AA\ for X1822--371 and between -0.484 \AA\ and 0.351 \AA\ for the comparison star.
These shifts were then applied to the spectra with the IRAF task SPECSHIFT.
Naturally this procedure does not take into account the possible non-linearity of the wavelength scale and its variation over night due to instrumental flexure, but these are considered to be very small effects.

Next the spectra were exported to ASCII format and further processing and analysis were done with the MOLLY software package developed by Tom Marsh.\footnote{http://deneb.astro.warwick.ac.uk/phsaap/software/}
In MOLLY the spectra were rebinned to the heliocentric frame, then normalised by fitting a fourth order spline to the continuum. 
The corresponding spline was subtracted from each spectrum.
Lastly, the resulting continuum-subtracted spectra were binned into 80 phase bins using the quadratic ephemeris by \citet{baptista}.
The number of spectra per phase bin varied from 1 to 4 (7 bins included only one spectrum).


\section{Analysis}

\subsection{Average spectrum and light curve}

   \begin{figure}
   \centering
\resizebox{\hsize}{!}{\includegraphics{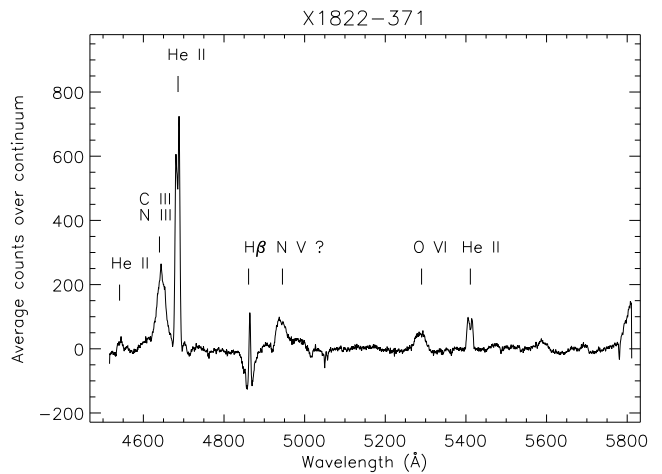}}
   \caption{Normalized average spectrum of X1822--371. The rising feature on the red end of the spectrum is an artefact from reductions at the edge of the CCD.
              }
         \label{FigOptAvSpec}
   \end{figure}
%

   \begin{figure}
   \centering
\resizebox{\hsize}{!}{\includegraphics{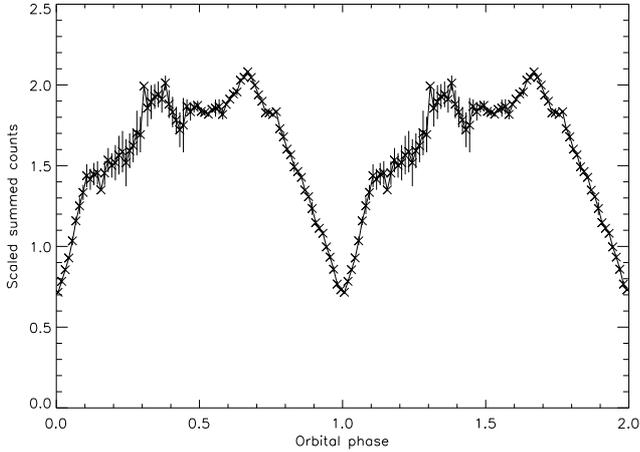}}
      \caption{Folded and binned light curve of X1822--371 in the wavelength range of 4560-5860 \AA\ integrated from individual spectra.
              }
         \label{FigOptLc}
   \end{figure}
%
   \begin{figure*}
   \centering
   \resizebox{\hsize}{!}{\includegraphics{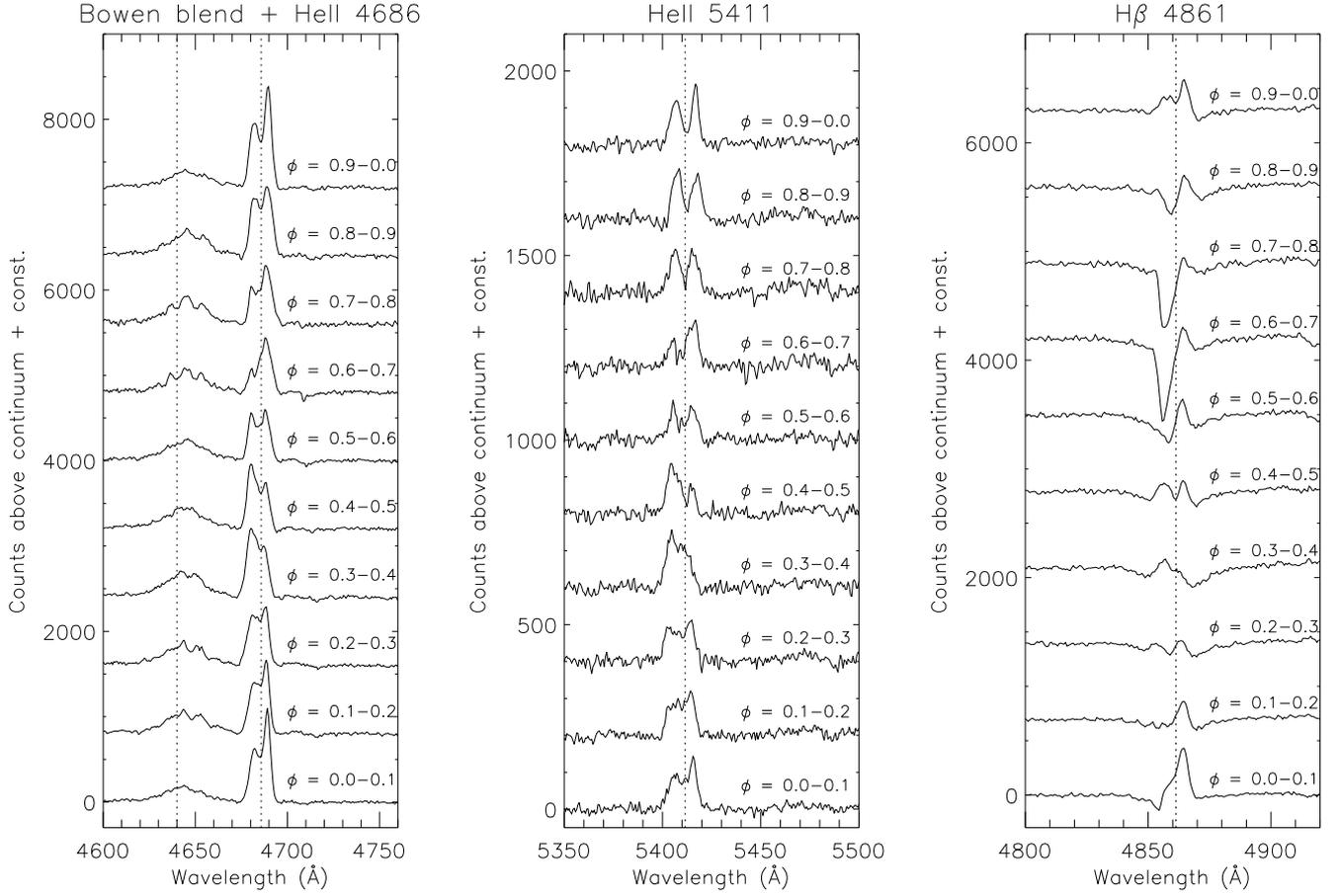}}
      \caption{Phase binned spectra (10 bins) over the orbital phase. The dotted lines mark the wavelengths of the lines (in case of the Bowen blend 4640 \AA.)
              }
         \label{FigSpecPh}
   \end{figure*}

The average spectrum of X1822--371 is shown in Figure \ref{FigOptAvSpec}.
The most prominent feature in the spectrum is the \ion{He}{ii} 4686 \AA\ emission line that shows a double-peaked profile typical for an accretion disc at high inclination.
Another, similar but weaker, \ion{He}{ii} line is seen at 5411 \AA.
A blend of \ion{C}{iii} and \ion{N}{iii} lines, the Bowen blend, is present in the 4640-50 \AA\ region.
H$\beta$ at 4861 \AA\ has a complex profile composed of superimposed absorption and emission components.
Other lines that can be seen in the average spectrum, which are however difficult or impossible to distinguish in individual spectra, we identify as follows.
The feature at the blue edge of our spectrum could be \ion{He}{ii} line at 4541 \AA\, detected also by \citet{cowley2003}, but not noted by \citet{casares}.
The bump on the red side of H$\beta$ could be a blend by \ion{N}{V} at 4945 \AA\ and the broad emission line at 5290 \AA\ is possibly caused by \ion{O}{vi}, as suggested by \citet{cowley2003}.

We also made a light curve by integrating the spectra.
The values in the light curve are just sums of the counts in individual spectra, divided by the corresponding values of the comparison star to correct for the slit losses and variable atmospheric conditions (this includes an assumption that the comparison star is not variable and that it is equally well placed on the slit all the time).
The errors of the light curve were estimated simply using photon noise.
Then the light curve was folded and binned into 80 phase bins.
The result is shown in Figure \ref{FigOptLc}.

The light curve shows similar structure to previous optical light curves.
The eclipse is broad and V-shaped and the ingress is slightly slower than the steeper egress.
The variability outside the eclipse is also similar to previously published light curves.
The slow rise after the eclipse is thought to be due to a thicker structure in the accretion disc.
However, when interpreting details of our light curve one should remember how it has been made and take into account the possible sources of error.

A light curve constructed from the wavelength range of 5052--5268 \AA\ where there are no discernible emission or absorption features has similar shape as the light curve in Fig. \ref{FigOptLc}.

\subsection{Spectral line variability}

The variability of the emission lines of the Bowen blend and \ion{He}{ii} 4686 \AA, \ion{He}{ii} 5411 \AA\ and H$\beta$ is shown in Fig. \ref{FigSpecPh} where each spectral feature is plotted at 10 different orbital phases.
As one can see, the Bowen blend is quite faint compared to \ion{He}{ii} 4686, however, there are some changes in its structure.
At phases 0.1 to 0.4 and 0.6 to 0.9 there are clearly two or three peaks (which we identify as 4634.12 \AA\ and 4640.64 \AA\ from \ion{N}{iii} and 4647.4 \AA\ from \ion{C}{iii}), with three peaks most prominent at phases 0.6 to 0.8.
The Bowen blend emission is thought to originate from the heated face of the secondary star \citep{cornelisse}.
The peaks are seen when the secondary is visible to us, that is before and after orbital phase 0.5.
This implies that around orbital phase 0.5 itself there is structure in the disc that is partially blocking our view of the donor star.

The \ion{He}{ii} lines at 4686 \AA\ and 5411 \AA\ show similar variability, even though it seems that in the 5411 \AA\ line the central trough between the peaks is deeper than in the 4686 \AA\ line.
More analysis of the Bowen blend and \ion{He}{ii} lines is presented below when Doppler tomography is discussed.

The H$\beta$ variability is more complex.
It appears that there is deep phase dependent absorption superimposed on a double peaked emission
The absorption feature seems to be sweeping/travelling through the emission from red to blue starting around orbital phase $\phi = 0.1-0.2$.
The absorption is deepest at phase 0.6--0.7, but it is not clear whether this is due to an increase in the absorption or a decrease in the emission component that is ``filling'' in the profile.

To study whether the emission lines also get eclipsed during the photometric eclipse we integrated the line fluxes (total counts above continuum, normalised with the continuum level of the comparison star spectrum to correct for atmospheric and other changes, errors were estimated again using the photon noise) and found that the \ion{He}{ii} 4686 \AA\ and 5411 \AA\ lines do not get eclipsed.

\begin{figure}
   \centering
\resizebox{\hsize}{!}{\includegraphics[]{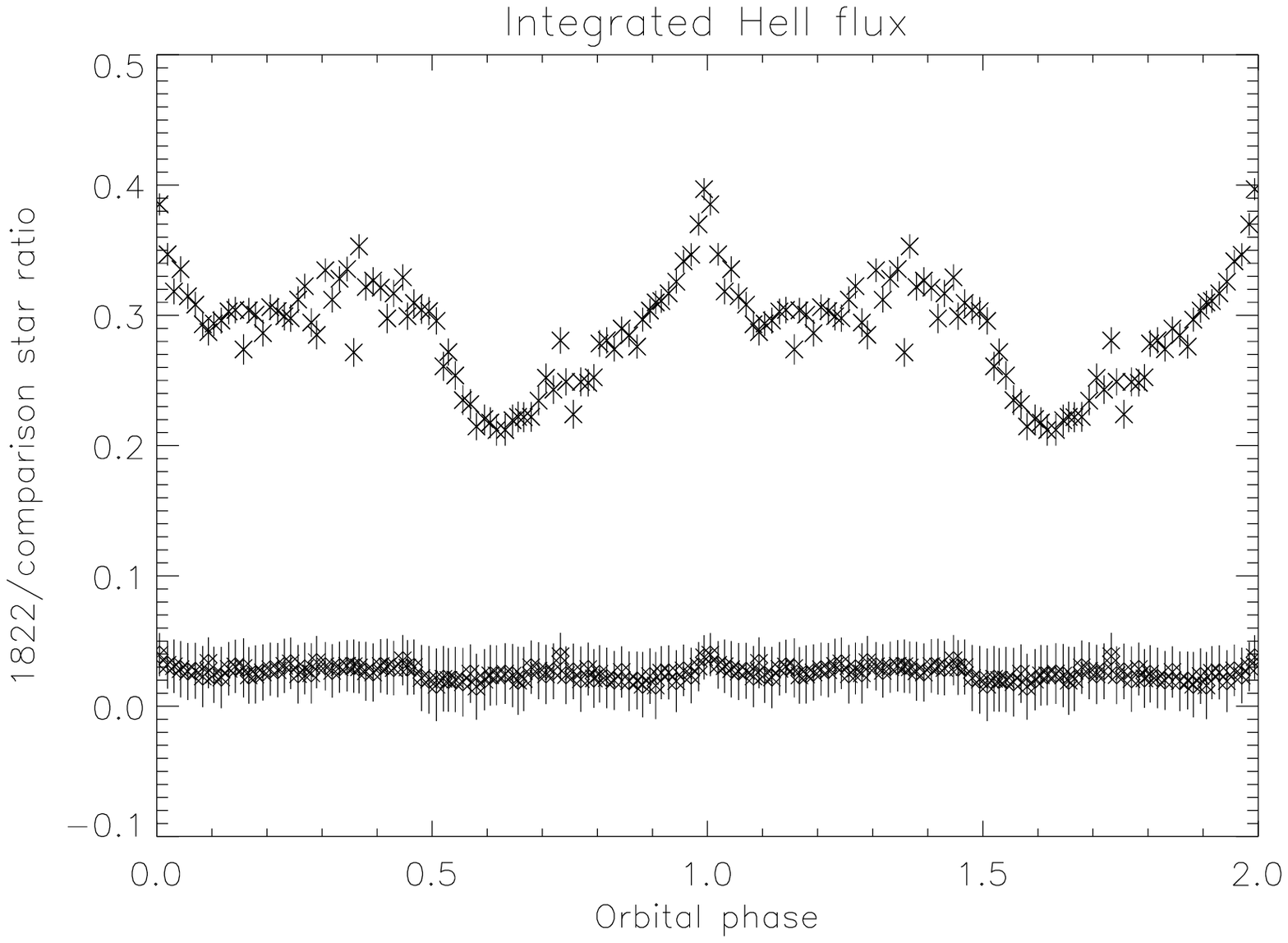}}\\
\resizebox{\hsize}{!}{\includegraphics[]{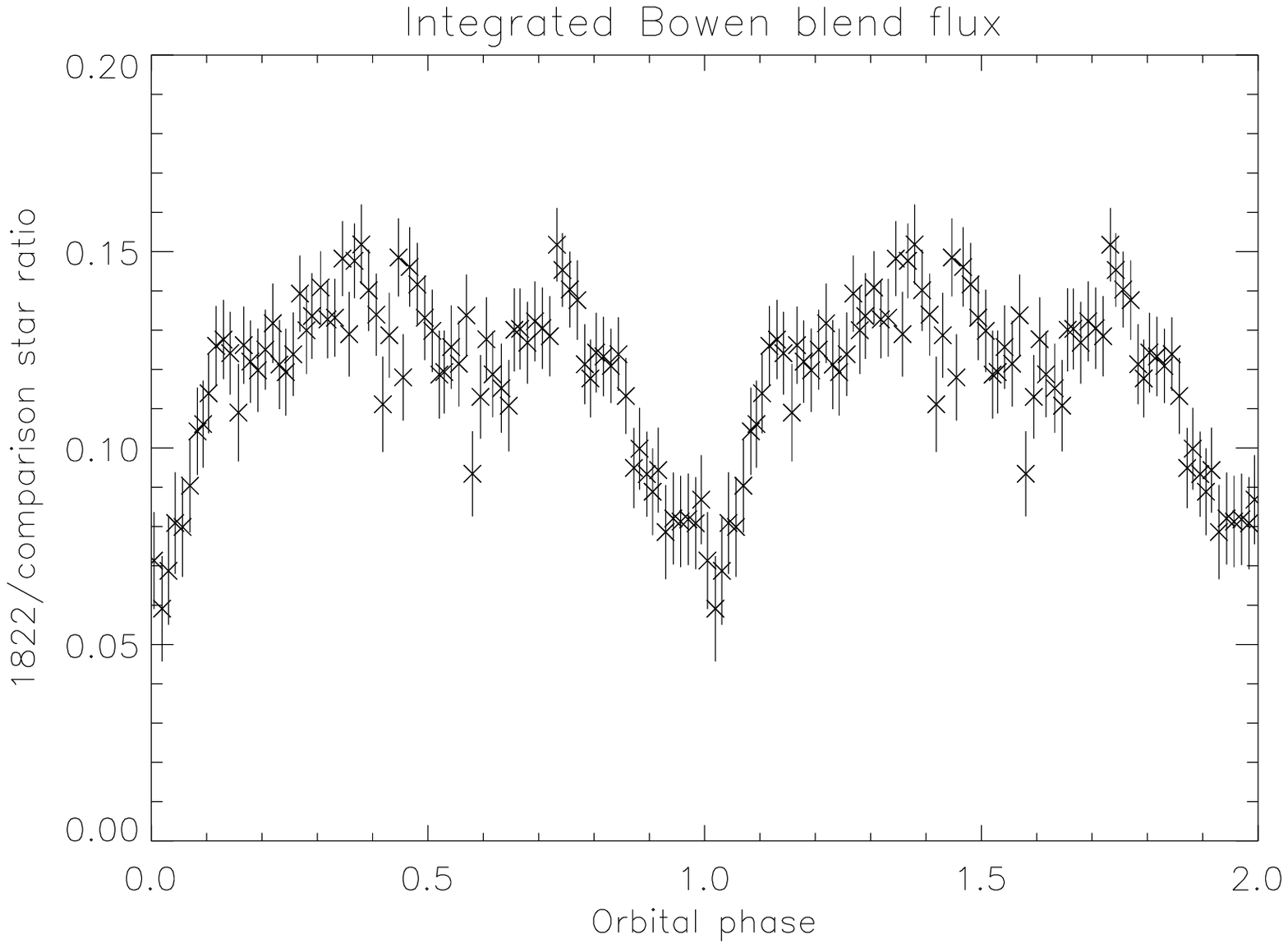}}
\caption{Integrated line flux of \ion{He}{ii} 4686 \AA\ and 5411 \AA\ (\textit{top}) and the Bowen blend (\textit{bottom}) normalised with the comparison star.}
\label{FigIntFlux}
\end{figure}

The integrated \ion{He}{ii} flux shows clear variability with the orbital phase, see Figure \ref{FigIntFlux}.
It brightens during the eclipse, which can be explained, at least partly, with the decreasing level of the continuum.

The Bowen blend flux does not vary significantly outside the eclipse (see Fig. \ref{FigIntFlux}) and actually it resembles the light curve.
This is well in accordance with the Bowen emission originating from the heated face of the secondary star.

We also checked the behaviour of the spectral feature of \ion{O}{vi} at 5290\AA\ to compare it to the result by \citet{hutchings}. 
They find that the \ion{O}{vi} emission line at 1032\AA\ shows variability over the orbital cycle.
Because the feature at 5290\AA\ is faint, we divied the orbital cycle to only 10 phase bins.
The integrated line flux shows variability with the orbital phase, but is not eclipsed and peaks at phase 0.45, where it is about 1.5 times brighter than at other phases.
The absence of eclipse suggests that the emission does not arise close to the compact object, contrary to what \citet{hutchings} suggest (they have only four data points and none at phase 0).

\subsection{Radial velocity curve}

In order to investigate the system parameters of X1822--371, we used the \ion{He}{ii} 4686 \AA\ line to make a radial velocity curve.
This was done by fitting two Gaussians to the emission line wings, as described by \citet{schneider}.
Because the emission line wings are thought to originate from the inner accretion disc, their motion will reflect the motion of the neutron star.
The width of the Gaussians was set to 100 km s$^{-1}$ and the separation of the two Gaussians, $a$, was varied between 400 and 1500 km s$^{-1}$ first with a step size of 100 km s$^{-1}$, later with a step size of 25 km s$^{-1}$ around the minimum.
Next the equation
\begin{equation} \label{Eq:radvel}
V = \gamma - K \sin (2 \pi (\phi - \phi_0))
\end{equation}
was fitted to the resulting radial velocity curves, where $V$ is the radial velocity, $\gamma$ is the systemic velocity, $K$ is the radial velocity semiamplitude and $\phi$ is the orbital phase and $\phi_0$ the orbital phase zero point.

\begin{figure}
   \centering
\resizebox{\hsize}{!}{\includegraphics{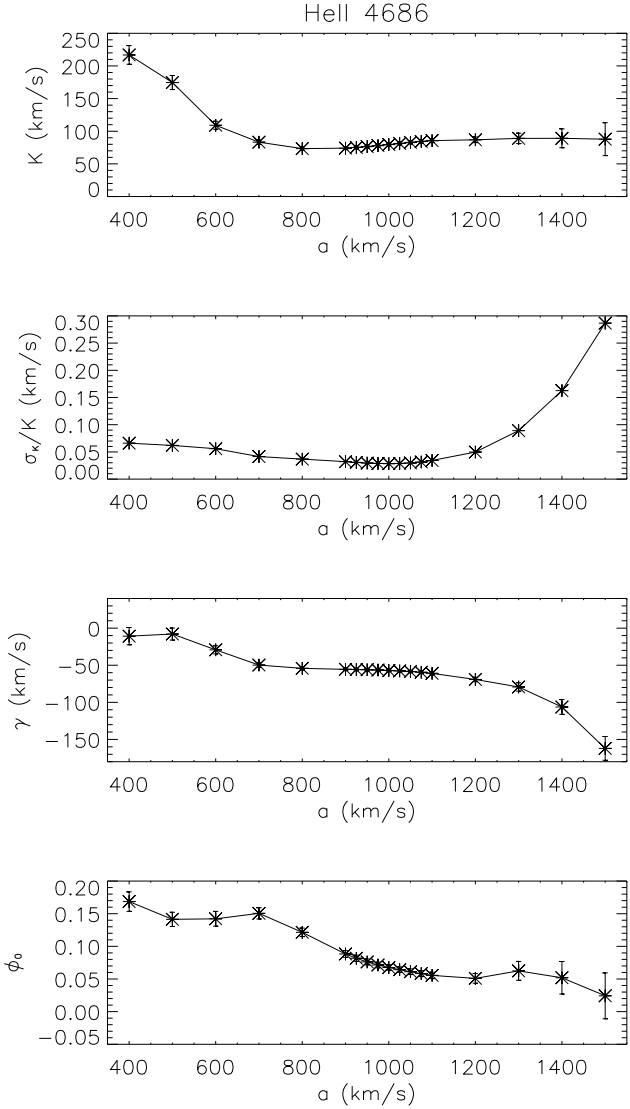}}
      \caption{Diagnostic diagram for \ion{He}{ii} 4686 \AA.
              }
         \label{Figdiag4686}
   \end{figure}

\begin{figure}
   \centering
\resizebox{\hsize}{!}{\includegraphics{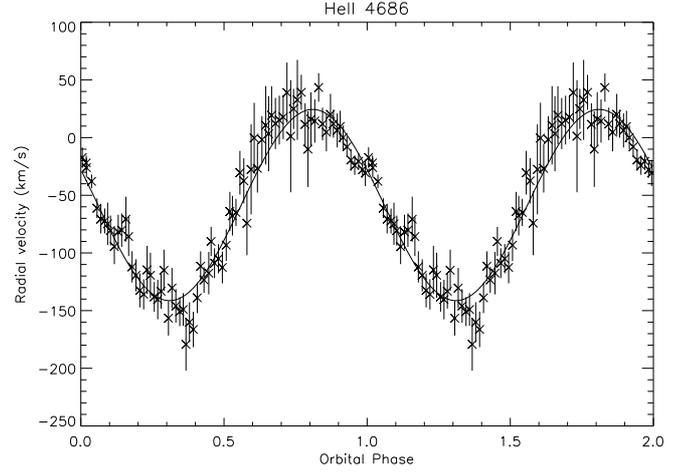}}
      \caption{Radial velocity curve of \ion{He}{ii} 4686 \AA\ emission line wings, likely reflecting the motion of the neutron star.
              }
         \label{Figradvel4686}
\end{figure}

The results of the fits are presented in a diagnostic diagram in Fig. \ref{Figdiag4686}.
The diagram is used to find the largest useful separation of the Gaussians as presented by \citet{shafter}.
As Figure \ref{Figdiag4686} shows, the minimum of $\sigma_K/K$ lies close to $a = 1000$ km s$^{-1}$.
We elected to use the value of $a =1050$ km s$^{-1}$ based on the lowest sum of squares of the fit.
With that separation of the Gaussians, the resulting fit parameters are: systemic velocity $\gamma = -58.5 \pm 1.7$ km s$^{-1}$, radial velocity semiamplitude $K = 82.9 \pm 2.5$ km s$^{-1}$, and the orbital phase zero point $\phi_0 = 0.06 \pm 0.004$.
The radial velocity curve (data and the fit) is presented in Fig. \ref{Figradvel4686}.
The results vary little with different values of $a$, e.g. with $a =950$ km s$^{-1}$: $\gamma = -56.1 \pm 1.6 $ km s$^{-1}$, $K = 76.6 \pm 2.3 $ km s$^{-1}$, and $\phi_0 = 0.76 \pm 0.004$, or with $a =1200$ km s$^{-1}$: $\gamma = -69.2 \pm 3.0  $ km s$^{-1}$, $K =  86.9 \pm 4.3  $ km s$^{-1}$, and $\phi_0 = 0.05 \pm  0.007 $.

These values are in line with previous studies by \citet{casares} ($\gamma = -43 \pm 4$ km s$^{-1}$, $K = 63 \pm 3$ km s$^{-1}$) and \citet{cowley2003} ($\gamma = -106 \pm 2$ km s$^{-1}$, $K = 77 \pm 4$ km s$^{-1}$).

\subsection{Doppler maps}

\begin{figure}
   \centering
\resizebox{\hsize}{!}{\includegraphics[angle=270]{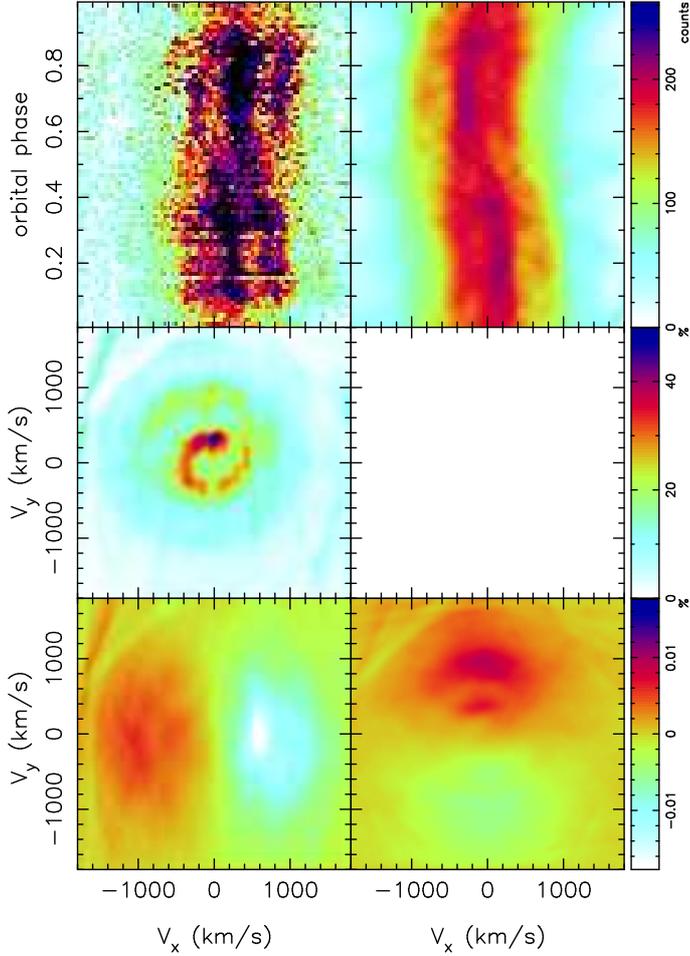}}
      \caption{Doppler map of the Bowen blend. Top left: observed data, top right: predicted data from the model, middle left: non-variable component / constant part of the map, middle right: amplitude of modulation, bottom left: cosine, bottom right: sine. The units of data and model are counts over continuum level, percentage for the modulation amplitude and its sine and cosine components.
              }
         \label{FigBowenMap}
   \end{figure}

\begin{figure}
   \centering
\resizebox{\hsize}{!}{\includegraphics[angle=270]{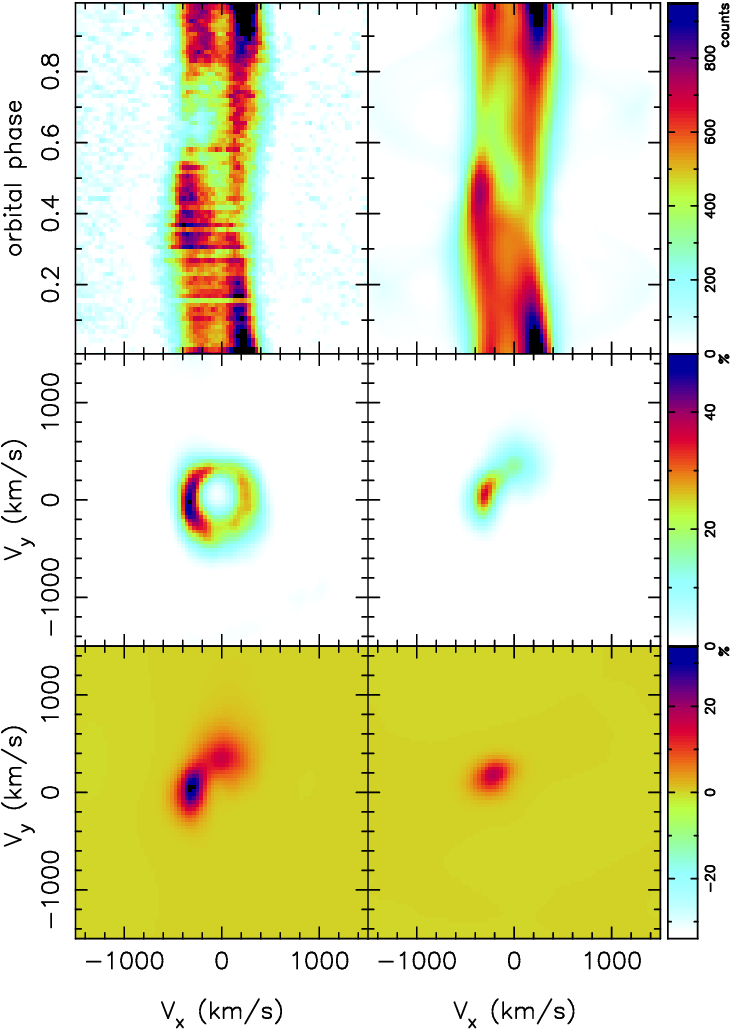}}
      \caption{Doppler map of \ion{He}{II} 4686 \AA. Top left: observed data, top right: predicted data from the model, middle left: non-variable component / constant part of the map, middle right: amplitude of modulation, bottom left: cosine, bottom right: sine. The units of data and model are counts over continuum level, percentage for the modulation amplitude and its sine and cosine components.
              }
         \label{FigHeII4686Map}
   \end{figure}

\begin{figure}
   \centering
\resizebox{\hsize}{!}{\includegraphics[angle=270]{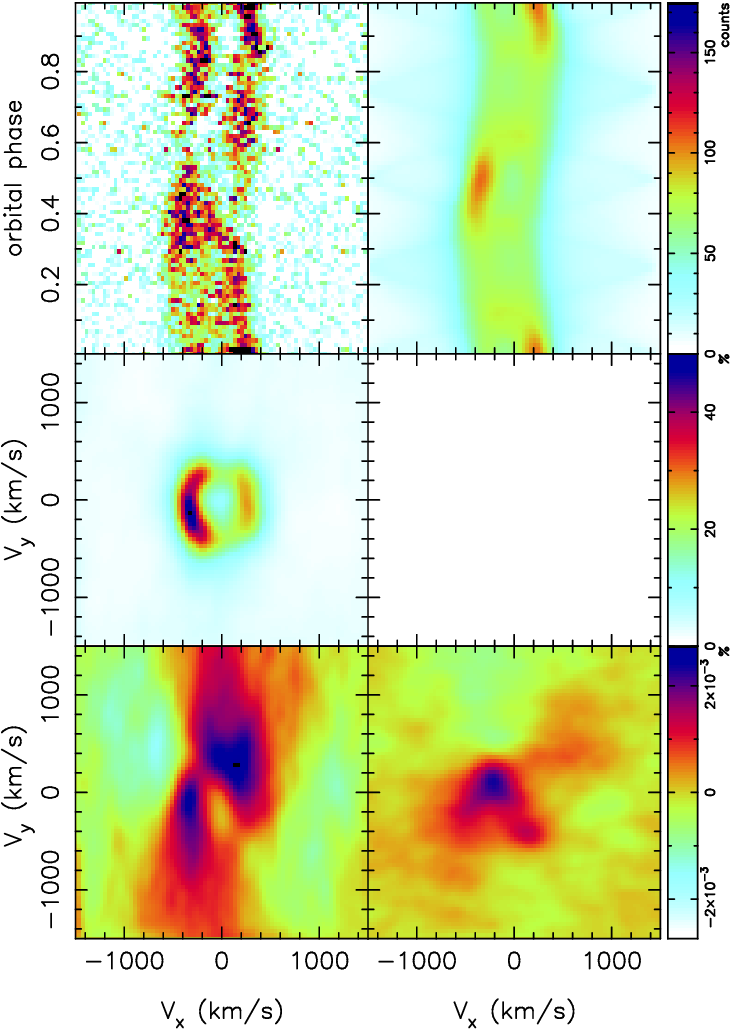}}
      \caption{Doppler map of \ion{He}{II} 5411 \AA. Top left: observed data, top right: predicted data from the model, middle left: non-variable component / constant part of the map, middle right: amplitude of modulation, bottom left: cosine, bottom right: sine. The units of data and model are counts over continuum level, percentage for the modulation amplitude and its sine and cosine components.
              }
         \label{FigHeII5411Map}
   \end{figure}

Although our view of X1822--371 does not fulfil all the axioms of Doppler tomography \citep{marsh2005}, for example the visibility of all elements of the disc does not remain constant in the case of a structured and thick accretion disc rim at high inclination, we applied this technique to our data, as previously done by \citet{harlaftis} and \citet{casares}.
We adopted the Doppler tomography code where the line flux is allowed to vary by \citet{steeghs2003}.
For the systemic velocity we used the value obtained from the radial velocity curve fitting (-58 km s$^{-1}$).
We produced Doppler tomograms from the Bowen blend (4640 \AA), \ion{He}{ii} 4686 \AA\ and 5411 \AA, shown in Fig. \ref{FigBowenMap}, \ref{FigHeII4686Map} and \ref{FigHeII5411Map}, respectively.

The Bowen blend average map (corresponding to the ``classical" Doppler map) shows a strong bright spot, ring-like structure, and a larger, more diffuse and fainter emission region.
The bright spot falls in the location of the secondary star ($V_x = 0, V_y \approx 300$ km s$^{-1}$), also seen in the Bowen blend map by \citet{casares}.
At sufficiently high spectral resolution, the Bowen blend has been found to be a powerful diagnostic of the motion of the donor star, as first reported by \citet{steeghs} for Sco X-1.
The ring in the map could be from the accretion disc.
There seems to be no modulation in the Bowen blend flux as there is nothing visible in the modulation amplitude map and the cosine and sine modulation maps show only variation at negligible levels.
In the Bowen blend map by \citet{casares} there is only a bright spot at the location of the secondary star, but we see also emission spread over larger velocities.
This could be due to the fact that we used the whole Bowen blend (centred at 4640 \AA) for reconstructing the map whereas \citet{casares} made their map with only the 4640 \AA\ feature of the blend.
The trailed spectrogram (top left panel in Fig. \ref{FigBowenMap} shows other components of the blend.

The constant-component \ion{He}{ii} maps, 4686 \AA\ and 5411 \AA\, have similar structure, but only 4686 \AA\ shows modulation.
This is at the level of 20\%  and is located in velocity space at the position of the expected accretion stream - accretion disc interaction/impact zone.
Curiously, \ion{He}{ii} 5411 \AA\ does not show this feature.

It is possible to see in Figures \ref{FigBowenMap}, \ref{FigHeII4686Map} and \ref{FigHeII5411Map} how the``S-wave" of the data has opposite phasing in the Bowen and \ion{He}{ii} maps.
This is just as we would expect if the Bowen emission originates on the heated face of the donor, and the \ion{He}{ii} emission comes from the accretion disc, i.e. the first traces the motion of the secondary and the latter that of the primary.

The \ion{He}{ii} emission seems to be peaking at around the velocity radius of 300 km s$^{-1}$ and the maximum velocity radius of significant emission is around 400 -- 600 km s$^{-1}$.
This is a relatively low velocity for matter in a neutron star LMXB accretion disc.

Thus we calculated what would be the theoretical observed radial velocity from the outer edge of a Keplerian accretion disc.
To estimate the minimum velocity a disc edge can have, we assume that the emitting material lies in the orbital plane and that its distance from the neutron star is the radius of the L1 point (a disc larger than that cannot exist).
Applying the system parameters of X1822--371 we derive an orbital velocity of $\sim$ 500 -- 600 km s$^{-1}$ (depending on the value of $M_1$). 
If the emission comes from the inner disc then the velocity should be larger.
These velocities are clearly larger than our data shows.
Thus this implies that the emission cannot originate in the orbital plane of the system - a fact also supported by the lack of eclipses in \ion{He}{ii} emission line flux.

%

%

%

\section{X-ray data}

We used archival X-ray data of X1822--371 obtained with the EPIC pn instrument onboard {\xmm}.
The observation was done on March 6, 2001 (Obs-Id 0111230101) and its duration was almost 52 ks covering approximately 2.5 orbital cycles of X1822--371.
We produced a light curve and phase-resolved spectra from the European Photon Imaging Camera (EPIC pn, \citet{struder}) data and high resolution spectra from both Reflection Grating Spectrometers (RGS, \citet{denherder}).

The data were processed using the {\xmm} Science Analysis System (SAS).
To extract the EPIC pn data we used a rectangular aperture centred on the source, since in timing mode the EPIC pn data has only one spatial coordinate \citep{struder} and the source appears as a vertical stripe in the image made from the event list.

\begin{figure}
   \centering
\resizebox{\hsize}{!}{\includegraphics[angle=0]{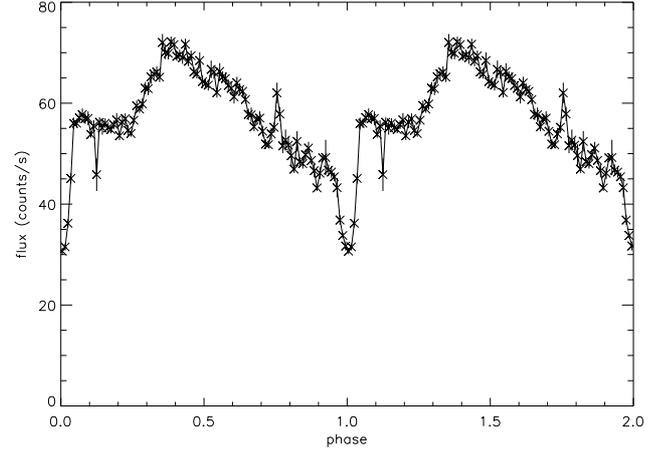}}
\caption{The folded and binned X-ray (0.2--10 keV) light curve of X1822--371.}
\label{FigXLc}
\end{figure}

The extracted X-ray light curve, covering the energy range of 0.2--10 keV, is shown in Fig. \ref{FigXLc}. The background subtraction was omitted since the source is relatively bright and our main goal was to study the spectra.

Next the phase-resolved spectra covering the energy range 0.3--10 keV were reduced with FLAG=0 and PATTERN=0-4.
The spectra were background-subtracted with background extracted from the data using the same aperture and parameters as for the source spectra, but the aperture was placed off the source.
The spectra were grouped with FTOOLS grppha task so that the number of counts in each group of channels was greater than 40.
We obtained altogether 10 spectra over the orbital period, each covering 0.1 in phase, except the eclipse spectrum which covers only 0.04 of the phase.
The orbital phases of the resulting spectra are listed in Table \ref{arvot}.

The RGS spectra, observed in the high event rate mode, were also reduced with SAS, using RGSPROC.
The spectra in the energy range of 0.3--2.8 keV were grouped with FTOOLS grppha task.
The spectra contain gaps due to poor calibration and damage in the detectors.

\begin{figure}
   \centering
\resizebox{\hsize}{!}{\includegraphics[width=8.8cm, angle=0]{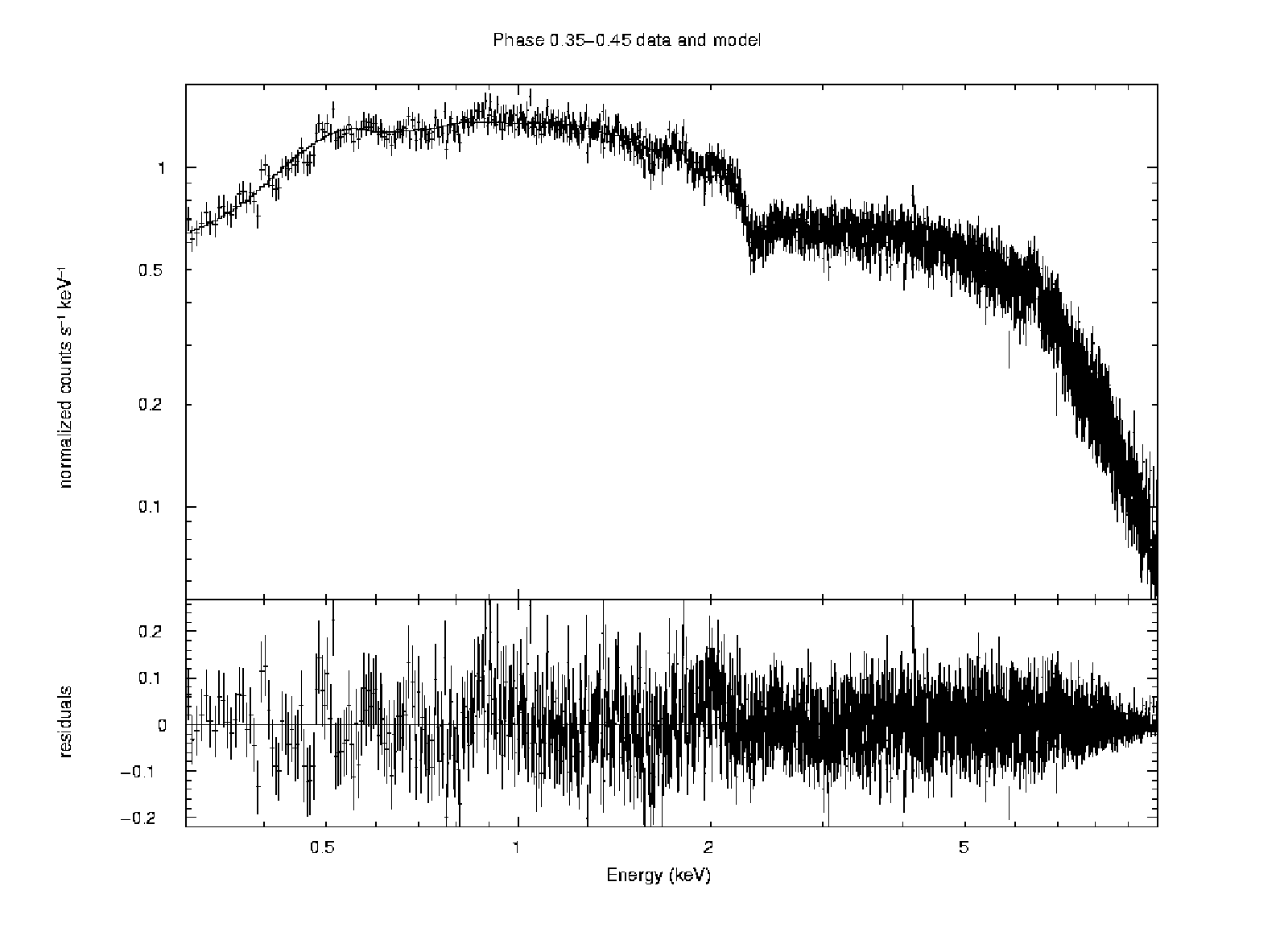}}
\resizebox{\hsize}{!}{\includegraphics[width=8.8cm, angle=0]{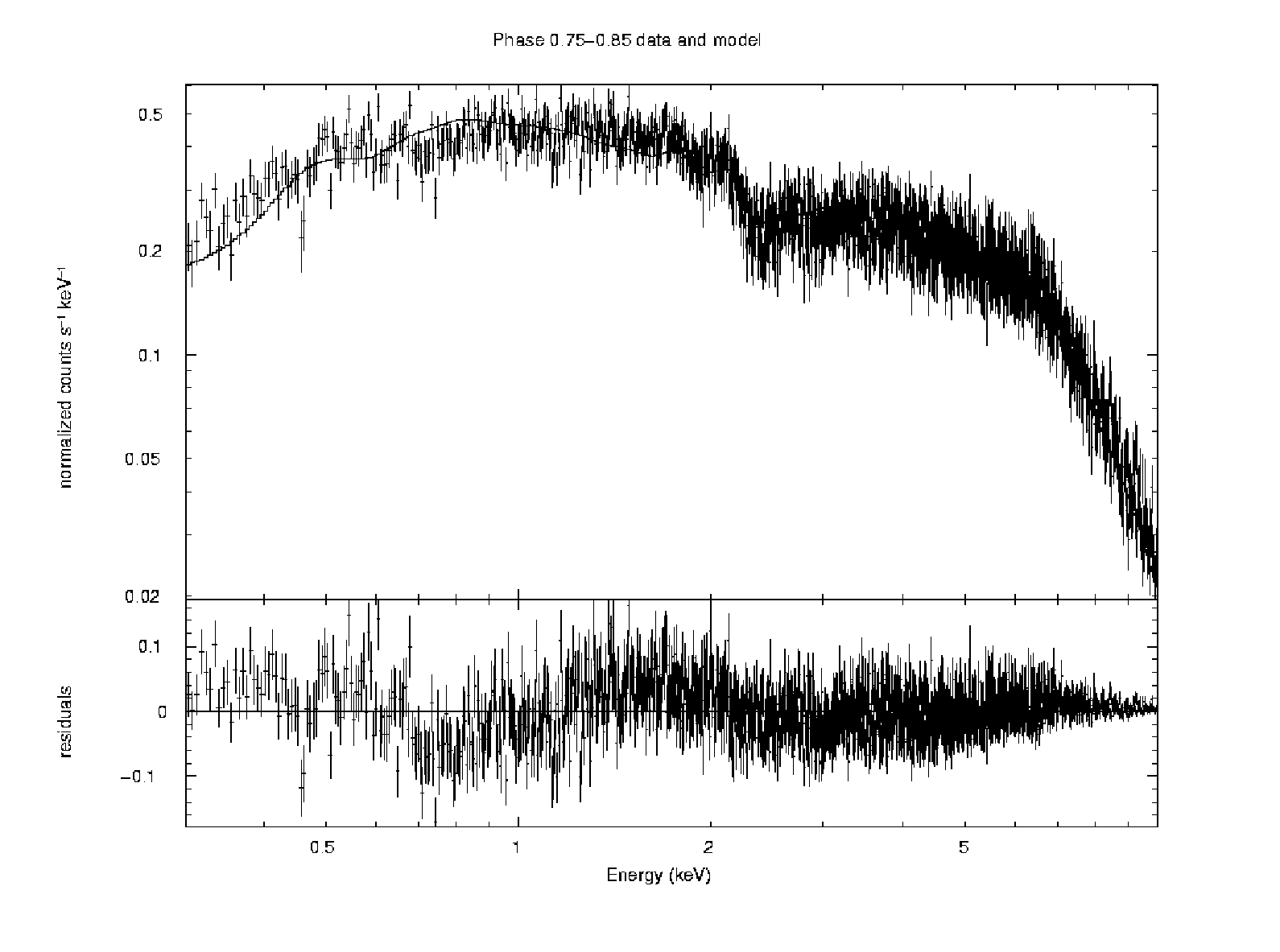}}
\caption{The 0.3--10 keV spectrum of X1822--371 at orbital phases 0.35--0.45 (\textit{top}) and 0.75--0.85 (\textit{bottom}).}
\label{FigXSpec1}
\end{figure}

The spectra were fitted with XSPEC v11.3 \citep{arnaud}.
Previous studies of X1822--371 with other X-ray observatories such as EXOSAT, BeppoSAX, RXTE and ASCA have shown that the spectrum is more complex to model than just a simple power law and/or blackbody with an emission line
profile (usually for the iron emission line at 6.4 keV).
The models of \citet{helliermason, parmar, heinz} and \citet{iaria} were tested for the {\xmm} data but they did not fit adequately.

Our best-fit model of the {\xmm} data is composed of four emission components and two absorption components.
These are multicolour black body model for an accretion disc (DISKBB), emitting hot diffuse gas model (MEKAL), Gaussian line profile (GAUSSIAN), a photon power law (PL), a photo-electric absorption model (WABS) and a partial covering fraction absorption (PCFABS), combined as follows:
\begin{displaymath}
\mbox{\small{WABS}} \cdot \mbox{\small{PCFABS}} \cdot (\mbox{\small{DISKBB}} + \mbox{\small{MEKAL}} + \mbox{\small{GAUSS}} + \mbox{\small{PL}})
\end{displaymath}
The physical justification for each component is the following.
The photo-electric absorption model accounts for the interstellar absorption in the line of sight to X1822--371.
The partial covering fraction mimics the accretion disc rim of X1822--371 which acts as an optically thick wall preventing a fraction of the emission from reaching us.
The emission components originate from the accretion disc or its immediate vicinity.
The disc black body is a superposition of multicolour black body elements, following the temperature gradient of an accretion disc.
The MEKAL model of a thermal plasma models the line emission from the accretion disc corona and the bremsstrahlung emanating from the accreting matter hitting the neutron star surface.
MEKAL was not able to fit the strong iron K$\alpha$ emission line at 6.4 keV, thus, a separate Gaussian line profile was required.
The non-thermal emission is represented with a power law model.
In the case of X1822--371 the spectrum includes a non-thermal hard tail which is due to photons Comptonized in the accretion disc corona.

First the model was fitted to the strongest spectrum  i.e. the spectrum covering phases 0.35--0.45.
The Gaussian line profile energy and its width were fixed to values of 6.4 and 0.1 keV, respectively.
Subsequently, after a fit to this first spectrum had been found, the values of the normalizations of the disc black body, MEKAL and power law were frozen and the model was fitted to the rest of the spectra, one spectrum at a time.

The reduced $\chi^2$ values for all the fits are reasonable and lie between 1.06 and 1.31.
The spectra together with the best-fit models at two different orbital phases are plotted in Figure \ref{FigXSpec1}.
The parameter values of the best-fit model are presented in Table \ref{arvot} which shows also the estimated errors corresponding to the 90\% confidence range.

\begin{figure}
   \centering
\resizebox{\hsize}{!}{\includegraphics[width=8.8cm, angle=0]{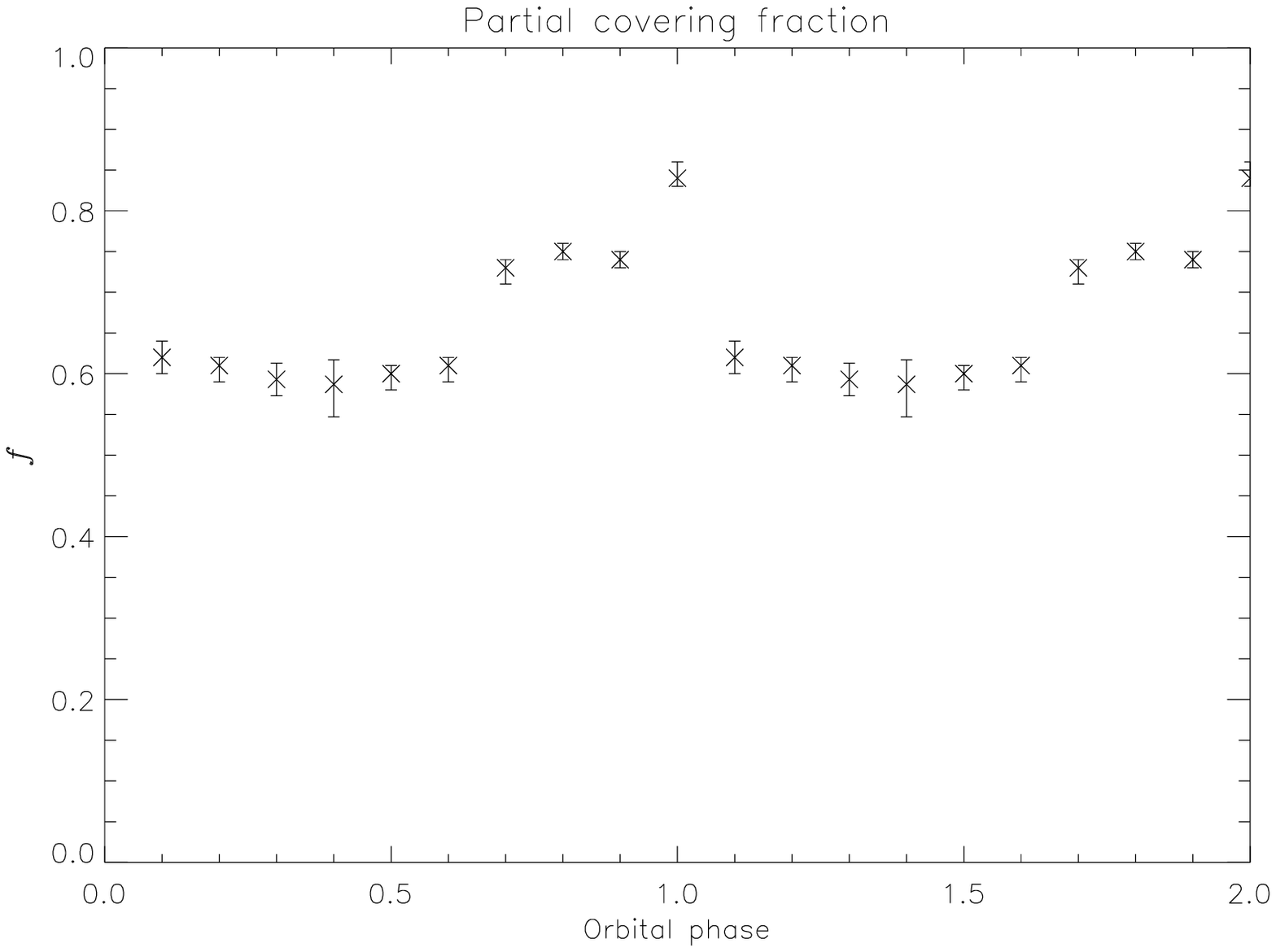}}
\caption{The partial covering fraction of the best-fit model over the orbital period.}
\label{FigCvrFract}
\end{figure}

The covering fraction orbital variation of the partial covering model is presented in Figure \ref{FigCvrFract}.
If the parameter mimics the vertical structure of the accretion disc it shows that the height of the disc increases at phases prior to the eclipse, that is from  orbital phase 0.6 to eclipse.

\begin{table*}
\caption{Best-fit model parameters and 90\% confidence errors for X1822--371 XMM spectra as a function of orbital phase.}
\begin{tabular}{l|llllllllll} 
\hline
Orbital Phase & 0.05-- & 0.15-- & 0.25-- & 0.35-- & 0.45-- & 0.55-- & 0.65-- & 0.75-- & 0.85-- & 0.98--\\
 & 0.15 & 0.25 & 0.35 & 0.45 & 0.55 & 0.65 & 0.75 & 0.85 & 0.95 & 0.02\\ 
\hline
$N_{H,wabs}$ ($\times 10^{22}$)  & 0.28$_{-0.02}^{+0.02}$& 0.29$_{-0.02}^{+0.02}$ & 0.30$_{-0.02}^{+0.02}$ & 0.27$_{-0.04}^{+0.05}$ & 0.25$_{-0.02}^{+0.02}$ & 0.26$_{-0.02}^{+0.02}$ & 0.38$_{-0.02}^{+0.02}$  & 0.39 $_{-0.01}^{+0.02}$& 0.41$_{-0.02}^{+0.02}$ & 0.51$_{-0.05}^{+0.07}$\\
 & & & & & & & & & & \\
$N_{H, pcfabs}$ ($\times 10^{22}$) & 3.91$_{-0.27}^{+0.26}$ & 3.92$_{-0.24}^{+0.23}$ & 4.73$_{-0.20}^{+0.21}$ & 4.27$_{-0.27}^{+0.28}$ & 4.38$_{-0.26}^{+0.27}$ & 4.27$_{-0.29}^{+0.28}$ & 3.45$_{-0.28}^{+0.24}$ & 3.6 $_{-0.33}^{+0.26}$& 4.44$_{-0.33}^{+0.29}$ & 3.26$_{-0.19}^{+0.59}$\\
 & & & & & & & & & & \\
$f$ & 0.62$_{-0.02}^{+0.02}$ & 0.61$_{-0.02}^{+0.01}$ & 0.59$_{-0.02}^{+0.02}$ & 0.59 $_{-0.04}^{+0.03}$& 0.60$_{-0.02}^{+0.01}$ & 0.61$_{-0.02}^{+0.01}$ & 0.73$_{-0.02}^{+0.01}$ & 0.75$_{-0.01}^{+0.01}$ & 0.74$_{-0.01}^{+0.01}$ & 0.84$_{-0.01}^{+0.02}$\\
 & & & & & & & & & & \\
$T_{in}$ (keV) & 4.90$_{-0.02}^{+0.02}$ & 4.92$_{-0.02}^{+0.02}$ & 5.30$_{-0.01}^{+0.01}$ & 5.41$_{-0.20}^{+0.24}$ & 5.19$_{-0.03}^{+0.02}$ & 5.12$_{-0.03}^{+0.02}$ & 4.04$_{-0.01}^{+0.01}$ & 3.97$_{-0.01}^{+0.01}$ & 3.90$_{-0.02}^{+0.02}$ & 2.87$_{-0.01}^{+0.02}$\\
 & & & & & & & & & & \\
$norm_{diskbb}$ $(\times10^{-3})$ & 4.54$_{-0.56}^{+0.54}$ & 4.54$_{-0.56}^{+0.54}$ & 4.54$_{-0.56}^{+0.54}$ & 4.54$_{-0.56}^{+0.54}$ & 4.54$_{-0.56}^{+0.54}$ & 4.54$_{-0.56}^{+0.54}$ & 4.54$_{-0.56}^{+0.54}$ & 4.54$_{-0.56}^{+0.54}$ & 4.54$_{-0.56}^{+0.54}$ & 4.54$_{-0.56}^{+0.54}$\\
 & & & & & & & & & & \\
$kT_{mekal}$ (keV) & 0.13$_{-0.01}^{+0.01}$ & 0.13$_{-0.01}^{+0.01}$ & 0.14$_{-0.01}^{+0.01}$ & 0.14$_{-0.01}^{+0.01}$ & 0.13$_{-0.01}^{+0.01}$ & 0.13$_{-0.01}^{+0.01}$ & 0.14$_{-0.01}^{+0.01}$ & 0.14$_{-0.01}^{+0.01}$ & 0.14$_{-0.01}^{+0.01}$ & 0.12$_{-0.02}^{+0.02}$\\
 & & & & & & & & & & \\
$norm_{mekal}$ $(\times10^{-3})$& 4.42$_{-2.37}^{+4.24}$ & 4.42$_{-2.37}^{+4.24}$ & 4.42$_{-2.37}^{+4.24}$ & 4.42$_{-2.37}^{+4.24}$ & 4.42$_{-2.37}^{+4.24}$ & 4.42$_{-2.37}^{+4.24}$ & 4.42$_{-2.37}^{+4.24}$ & 4.42$_{-2.37}^{+4.24}$ & 4.42$_{-2.37}^{+4.24}$ & 4.42$_{-2.37}^{+4.24}$\\
 & & & & & & & & & & \\
$E_{Gauss}$ (keV) & 6.4 & 6.4 & 6.4 & 6.4 & 6.4 & 6.4 & 6.4 & 6.4 & 6.4 & 6.4\\
 & & & & & & & & & & \\
$\sigma_{gauss}$ (keV) & 0.1 & 0.1 & 0.1 & 0.1 & 0.1 & 0.1 & 0.1 & 0.1 & 0.1 & 0.1\\
 & & & & & & & & & & \\
$norm_{gauss}$ $(\times10^{-5})$& 1.56$_{-0.22}^{+0.60}$ & 2.55$_{-0.27}^{+0.52}$ & 2.93$_{-0.33}^{+0.52}$ & 2.87$_{-0.43}^{+0.42}$ & 3.11$_{-0.65}^{+0.26}$ & 2.49$_{-0.22}^{+0.73}$ & 1.39$_{-0.38}^{+0.20}$ & 0.98$_{-0.30}^{+0.22}$ & 1.17$_{-0.29}^{+0.23}$ & 0.0$_{-0.18}^{+0.16}$\\
 & & & & & & & & & & \\
$\alpha_{pl}$ & 2.96$_{-0.16}^{+0.20}$ & 3.11$_{-0.16}^{+0.20}$ & 3.30$_{-0.19}^{+0.22}$ & 2.98$_{-0.37}^{+0.43}$ & 2.75$_{-0.17}^{+0.19}$ & 2.80$_{-0.17}^{+0.20}$ & 3.64$_{-0.17}^{+0.25}$ & 3.61$_{-0.17}^{+0.26}$ & 3.68$_{-0.16}^{+0.23}$ & 4.61$_{-0.45}^{+0.21}$\\
 & & & & & & & & & & \\
$norm_{pl}$ $(\times10^{-3})$& 3.08$_{-0.52}^{+0.62}$ & 3.08$_{-0.52}^{+0.62}$ & 3.08$_{-0.52}^{+0.62}$ & 3.08$_{-0.52}^{+0.62}$ & 3.08$_{-0.52}^{+0.62}$ & 3.08$_{-0.52}^{+0.62}$ & 3.08$_{-0.52}^{+0.62}$ & 3.08$_{-0.52}^{+0.62}$ & 3.08$_{-0.52}^{+0.62}$ & 3.08$_{-0.52}^{+0.62}$\\
 & & & & & & & & & & \\
$\chi^{2}_{\nu}$ & 1.06 & 1.06 & 1.08 & 1.06 & 1.07 & 1.10 & 1.17 & 1.31 & 1.29 & 1.30\\
\end{tabular}
\label{arvot}
\end{table*}

\begin{figure}
   \centering
\resizebox{\hsize}{!}{\includegraphics[angle=0]{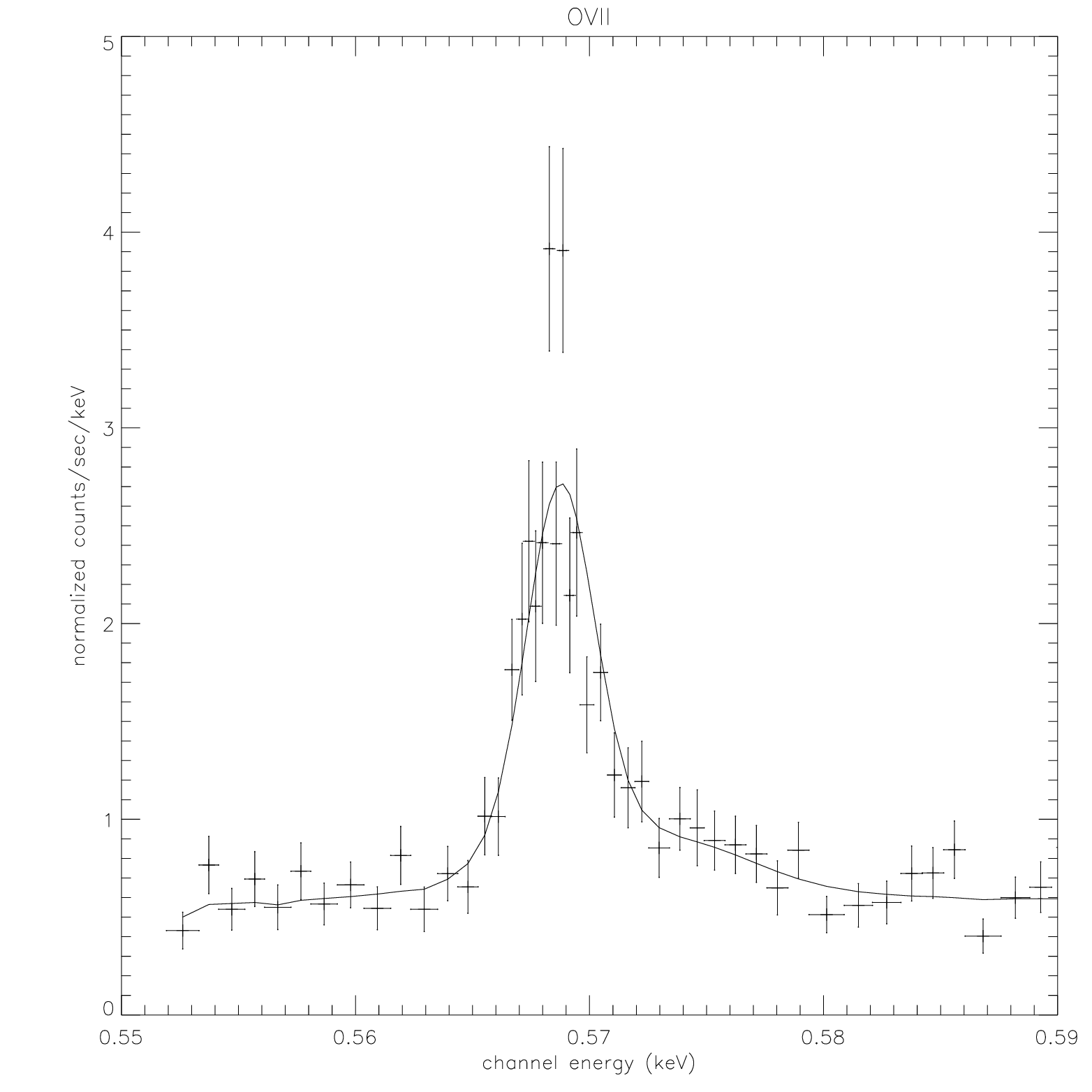}}
\caption{The \ion{O}{vii} line at 0.5685 keV, composed of the intercombination line at 0.5687 keV and resonance line at 0.574 keV shown together with the fit of the Gaussian line profiles.}
\label{FigOVII}
\end{figure}

Using \citet{zombeck} and \citet{jimenez-garate} we identified two emission lines in the RGS1 first order spectrum.
The presence of the lines was confirmed by fitting a Gaussian line profile at the position of the proposed lines, which are \ion{He}{}-like \ion{O}{vii} at 0.5685 keV and Ly$\alpha$-like \ion{O}{viii} at 0.654 keV.
The \ion{O}{vii} line is actually a complex composed of a stronger intercombination line at 0.5687 keV and a weaker resonance line at 0.574 keV.
The line components were fit with Gaussian profiles and no forbidden component at 0.561 keV was found (see Fig. \ref{FigOVII}).

As presented by \citet{porquet} the line ratios of \ion{He}{}-like ions can be used as plasma diagnostics.
The ionization process of the plasma can be determined by calculating the ratio
\begin{displaymath}
G = \frac{f + i}{r}
\end{displaymath}
where $f, i$ and $r$ stand for the relative intensities of forbidden, intercombination and resonance lines, respectively.
To calculate this ratio we used the equivalent widths from the Gaussian fits (see Table \ref{eqwidths}).
This way we obtained the value $G = 9.2$ which suggests that the emitting plasma is purely photoionized ($G > 4$).
As we detect no forbidden line, its ratio to the intercombination line is zero.
This fact combined with Figure 8 and 9 in \citet{porquet} infer that the electron density of the accretion disc corona is high, of order $ n_e = 10^{12}~ \mathrm{cm}^{-3}$.
Figure 11 in \citet{porquet} shows that the plasma is purely photoionized (the relative strengths of the lines are correct for purely photoionized and $n_e=10^{12}\mathrm{cm}^{-3}$, similar to our line profile).

\begin{table}
\caption{The measured equivalent widths of the forbidden (f), intercombination (i) and resonance (r) lines of OVII in the first order RGS1 spectrum.}
\begin{center}
\begin{tabular}{lll}
\hline
Line&Energy&EW \\
 &(keV)&(eV)\\
\hline
f&0.561&-\\
i&0.56875&14.1\\
r&0.574&1.53\\
\hline
\end{tabular}
\end{center}
\label{eqwidths}
\end{table}

\citet{ji} used Chandra data of X1822--371 and their analysis of the \ion{O}{vii} line ratio yields similar diagnostics as ours.


\section{Discussion and Conclusions}

In this paper we have presented phase-resolved optical and X-ray spectra of X1822--371.

The optical and X-ray light curves (Fig. \ref{FigOptLc} and \ref{FigXLc}) both show the secondary star contributing to the extended partial eclipse.
The eclipse is narrower in X-rays than in optical: in X-rays it lasts approximately 0.1 in orbital phase whilst in optical it lasts at least twice that.
This can be understood with the difference in the emission region size.
The out-of-eclipse variability is thought to be due to the varying height of the accretion disc rim.
In both curves there is a plateau-like section immediately following the eclipse egress.
This can be interpreted as an accretion disc rim that has some vertical structure covering the brighter inner disc which is opposite to the stream impact region, thereby producing maximum brightness around phase 0.5.

The semiamplitude of our \ion{He}{ii} 4686 \AA\ radial velocity curve is $K_1 = 82.9\pm 2.5$ km s$^{-1}$ which traces the motion of the NS.
If we interpret the bright spot on the Bowen blend map as a signature of the companion star its velocity is approximately $K_2 = 300$ km s$^{-1}$.
Thus, we can estimate the mass ratio of the components in the system to be $q = K_1/K_2 = 0.28$.
With a canonical NS mass of 1.4 M$_{\sun}$ the companion mass would be 0.39 M$_{\sun}$. But if the NS mass is as high as 2.32 M$_{\sun}$ \citep{munoz}, then the companion mass would be 0.64 M$_{\sun}$.

The Doppler map of \ion{He}{ii} shows low velocities of the emission line peaks implying that the emission does not come from the orbital plane of the system (assuming a Keplerian disc).
In addition, the \ion{He}{ii} flux is actually not eclipsed even if the continuum light is.
This could be explained with a model where the emission does not originate from the accretion disc itself, but from matter above the accretion disc, for example a wind, that still retains some of the velocity field of the disc.
Actually recent papers by \citet{bayless} and \citet{burderi} suggest the presence of a wind or some kind of outflow from the system.

The Doppler maps of \ion{He}{ii} 4686 \AA\ and 5411 \AA\ look similar, however, there is no variable component in the 5411 \AA\ map.
This is somewhat peculiar, as it would be expected that these lines should originate in roughly the same regions, even though they arise from different transitions.
The \ion{He}{ii} emission seems to be stronger at negative x-velocities, which corresponds to the area where the accretion stream from the secondary star hits the accretion disc and gets heated.

An intriguing model that can explain our data is that for CAL 87 by \citet{schandl}.
CAL 87 is a close binary super-soft source system in the Large Magellanic Cloud and it is considered to contain an ADC and a similar thick accretion disc as in X1822--371.
In this model a ``spray'' of matter is created in the location where the accretion stream hits the disc.
Hydrodynamical simulations by \citet{armitage} result in a similar model for the accretion stream impact region in the case of inefficient cooling.

The partial covering fraction of the best-fit X-ray model is higher prior to the eclipse than at other orbital phases (excluding the eclipse).
This implies there is an increased level of absorbing material in our line of sight, which could be the result of the``spray''.
Also the H$\beta$ feature in the optical spectra exhibits the strongest absorption feature at those phases.

The complex profile of H$\beta$ could be composed of two absorption components, one narrow and one wide, on which a double-peaked emission component is superimposed.
The double-peaked component would resemble the \ion{He}{ii} profile, originating from matter in a circular motion around the NS, in the orbital plane or above it.
The narrow absorption is caused by the ``spraying'' matter.
The absorption is strongest at orbital phases 0.6-0.8 when the spray would be in our line of sight to the emitting central parts of the accretion disc.
If the spray continues moving downstream after the accretion stream impact region, it could cause a redshifted feature at phases 0.8-1.
This absorption component is blueshifted (see Fig. \ref{FigSpecPh}), so it means that matter is being ``sprayed'' towards us, i.e. away from the accretion disc.
However, in addition to the spray away from the disc, a spray or stream over the disc is required to explain the redshifted smaller absorption feature after orbital phase 0.1.

The spray model is also supported by other studies, e.g. \citet{ji} observe some X-ray line emission to be located at the stream impact region and \citet{harlaftis} found \ion{Fe}{ii} $\lambda$6516 in absorption present only at the orbital phase 0.8.

In connection with the fact that the \ion{He}{ii} emission would arise from above the orbital plane, we calculated at what height should the emission originate in order to be visible throughout the whole orbital cycle.
With an inclination of 85$\degr$ the emission should come from a minimum height $z = 0.2  a$, where $a$ is the distance between the components.
If the inclination was 80$\degr$ the minimum height of $z = 0.15  a$ is sufficient so that the emission region is seen at all phases.

Another possibility is that the inclination angle is lower than has been modelled previously.
It would easily explain the observed lower velocities and the fact that the \ion{He}{ii} lines do not get eclipsed.
However, it would be in contradiction to the eclipses seen in the light curves.

Our multiwavelength study of the accretion disc in X1822--371 strengthens the case for the existence of non-axisymmetric vertical disc structure (``flared disc'') in interpreting the observed properties. 
This is the case for both the optical and X-ray data and these results agree with the optical and X-ray results for another ADC source, UW CrB \citep{hakala2005, hakala2009}. 
 The actual nature and cause of this vertical structure is not yet properly understood.

\begin{acknowledgements}
We thank Tom Marsh for the use of MOLLY
and Danny Steeghs for the use of MODMAP code.
A.S. acknowledges the financial support of the Finnish Graduate School in Astronomy and Space Physics.

\end{acknowledgements}

\bibliographystyle{aa} 
\bibliography{aa18439-11} 
\end{document}